\documentclass[11pt,twoside]{article}
\usepackage{cozumel2005}
\usepackage{epsf}
\usepackage{psfig}
\usepackage{graphicx}
\usepackage{epsfig}
\begin{document}
\title{The underlying stellar absorption contribution to the primordial helium abundance determination}    
\author{F. F. Rosales-Ortega, R. Terlevich, E. Bertone and M. Chavez}   
\affil{{\sc Instituto Nacional de Astrof\'{i}sica, \'{O}ptica y Electr\'{o}nica}
Apartado postal 51 y 216, CP 72000, Puebla, Pue. Mexico}    

\begin{abstract} 
We carried out an exploratory analysis of the contribution of the underlying stellar
absorption to the total uncertainty of the abundance of primordial helium using simple
stellar populations models and observational data from the Sloan Digital Sky Survey. Results indicate that our analysis yields a lower limit to the error on the helium abundance determination if the stellar absorption is neglected.
\end{abstract}


\section{Introduction}                      

The determination of $Y_p$, the mass fraction of primordial helium, is of
fundamental importance due to its important cosmological implications. (see
\citet{oli04} for a discussion). The discrepancy among di\-ffe\-rent
measurements of $Y_p$ has been reduced to a 1-2\% level, nevertheless this
uncertainty still imply a wide range of possible va\-lues of the
barion-to-photon ratio $\eta$, an important parameter of the Standard Big Bang
Nucleosynthesis, usually obtained from observed primordial abundances. The
classical method for the determination of $Y_p$ \citep{peim74} is based on the
abundance analysis of extra-galactic H II regions through their emission lines
and the extrapolation to zero metallicity.

Several systematic errors can affect a precise determination of the $Y_p$.
We focus here on the uncertainty produced by an underlying stellar absorption,
which causes a systematic decrement of the intensity of helium nebular
emission lines. An accurate assessment of this factor is important if one
requires high precision results.
 
Theoretical single stellar population (SSP) models appear as an effective tool
to perform a thorough investigation of the absorption contribution for the
lines commonly used to determine $Y_p$ such as He I $\lambda$4471,
$\lambda$5876, $\lambda$6678 and He II $\lambda$4686, which are nebular lines
of particularly small equivalent width. We present here the results of an
exploratory analysis on this issue.


\section{He stellar absorption}

We measured the equivalent width of the four He lines of interest (He I
$\lambda$4471, $\lambda$5876, $\lambda$6678, and He II $\lambda$4686) using
the pure absorption high resolution SSP models (Bressan 2005, private
communication) based on the theoretical libraries of stellar spectral called
{\bf BLUERED} and {\bf UVBLUE} \citep{bert04,rodri05} respectively.

In order to account for the possible presence of mixed population we
calculated 24 combinations of SSPs of different ages and chemical
composition. For these combinations we considered two metallicities
($Z_{\odot}/50$ and $Z_{\odot}/5$) and a set of {\em young} (10-500 Myr) and
{\em old} (1-2 Gyr) populations. The SED mixed populations ($F_{ij}$) were
calculated such that $F_{ij} = aF_{young_i}+bF_{old_j}$ where $a$ and $b$ were
0.8 and 0.2 respectively. The EW of He lines for each flux combination were
measured after the SEDs had been degraded to the spectral resolution of the
Sloan Digital Sky Survey ($R= \lambda/\Delta\lambda =1800$). Some examples of
SEDs (at a fixed age of 1 Gyr of the {\em old} component) in the region around
the He I $\lambda$4471 line are shown in Figure~1.

The intensity of the line falls abruptly with increasing age of the {\em
  young} population and its absorption becomes almost negligible for ages
older than 200 Myr. This same behavior is common to all the lines that we have
included in our analysis.

\begin{figure}[!t]
\centering
\includegraphics[height=10cm,angle=90]{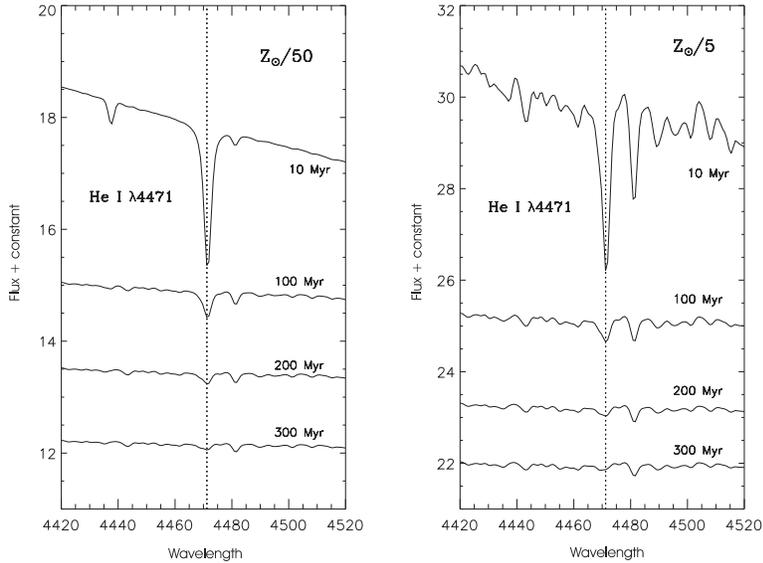}
\caption{Spectral energy distributions around the He I $\lambda$4471 line of
  SSPs with variable age of the {\em young} component and a fixed age of 1 Gyr
  of the {\em old} component, at metallicities of $Z_{\odot}/50$ ({\em left}
  panel) and $Z_{\odot}/5$ ({\em right} panel).} 
\end{figure}

\section{Comparison to observations}

In order to quantitatively asses the importance of the stellar absorption, we
measured the variation of the He abundance on the metal poor extragalactic H
II region SDSS J094401.86-003832.1, which we selected from the SDSS catalog
based on the quality observed spectrum. Figure 2 shows the observed spectrum
around the He I $\lambda$4471 emission line together with the theoretical SSP
flux with ages of 10 Myr and 1 Gyr for the {\em young} and {\em old}
components, respectively. This combination presents the largest stellar
absorption and we can notice that, even though it is small compared with the
emission feature, its effect on the EW is not negligible.
Table 1 reports the equivalent widths of the three He I lines in emission
(measured on the observed spectrum) and in absorption (measured on the SSP SED
partially shown in Figure 2 .) as well as the variations in the He ionic
abundance $y^+$ with and without taking into account the effect of underlying
stellar component. The He ionic abundance has been computed following
\citet{pagel92}. We found the maximum He abundance variation to be about 6\%
for the He\,I $\lambda$4471 line.

\begin{figure}[!t]
\centering
\includegraphics[height=7cm,angle=90]{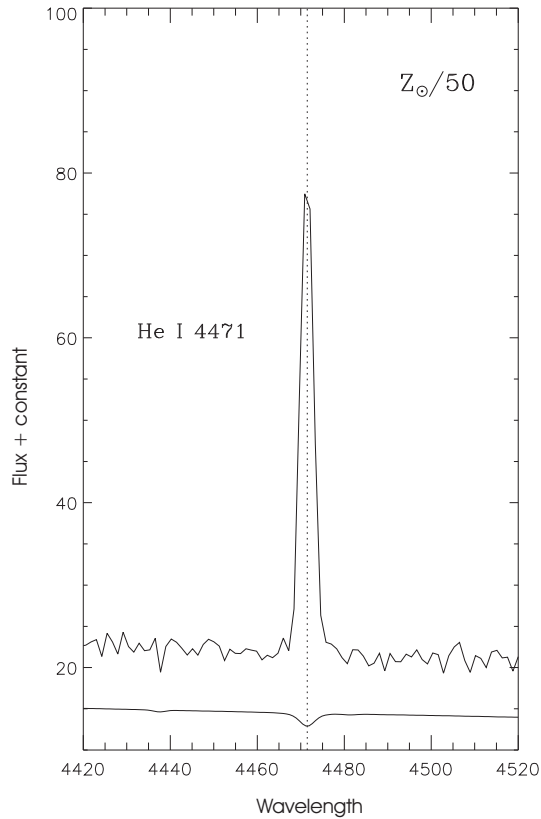}
\caption{SDSS J094401.86-003832.1 spectrum and theoretical SSP model with $Z_{\odot}/50$, $f_{young}$ = 10 Myr and $f_{old}$ = 1.5 x $10^9$ yr showing the influence of the absorption feature for He I $\lambda$4471.}
\end{figure}

\begin{table}[!ht]
\caption{The equivalent width of the He\,I lines and the He ionic abundance.}
\begin{center}
{\small
\begin{tabular}{ccccccc}
\tableline
\noalign{\smallskip}
& \multicolumn{2}{c}{Equivalent width (\AA)}  &&  \multicolumn{2}{c}{Abundance $y^+$ (x 10$^3$)}\\
\noalign{\smallskip}
\cline{2-3} && \cline{3-4} 
\noalign{\smallskip}
He {\footnotesize I}& Object & SSP model && {\footnotesize without correction} & {\footnotesize with correction} & $\Delta y^+$\\
\tableline
\noalign{\smallskip}
$\lambda$\,4471 & 9.04 $\pm$ 0.04 & 0.51 $\pm$ 0.02 && 87.02 $\pm$ 1.51 & 92.03 $\pm$ 1.50 & 5.1 (5.86\%)\\
$\lambda$\,5876 & 51.20 $\pm$ 0.3 & 0.36 $\pm$ 0.02 && 86.85 $\pm$ 1.45 & 87.40 $\pm$ 1.42 & 0.55 (0.60\%)\\
$\lambda$\,6678 & 18.4 $\pm$ 0.5  & 0.23 $\pm$ 0.01 && 81.78 $\pm$ 1.62 & 82.94 $\pm$ 1.65 & 1.16 (1.42\%)\\
\noalign{\smallskip}
\tableline
\end{tabular}
}
\end{center}
\end{table}

\section{Conclusions}

In the exploratory analysis presented here, we found (based on Table 1) that
the difference between the He abundance determined with and without taking
into account the underlying stellar absorption component is on average
3\%. Despite that this is a small value, the effect is not negligible, as such
a variation overestimate systematically the helium abundance of individual
regions which are then used to fit a linear regression in order to get a value
of $Y_P$. Greater values of $y^+$ would yield an overestimated primordial
helium abundance, therefore it is important to obtain an accurate correction
for this effect.

However, we stress that the grid of SSP models that we used had a lower age
limit of 10 Myr; younger stellar populations, with ages characteristic of O-B
stars, the typical ionizing sources of H\,II regions, should be taken into
account to better map their contribution to the EW of He lines. Therefore our
result represents a lower limit to the error on the helium abundance
determination if the stellar absorption is neglected.

\acknowledgements             

FFRO would like to thank the INAOE Astrophysics Department and specially
Miguel Chavez Dagostino for his kindly support which made him possible attend
such a successful con\-fe\-rence.



\begin{thebibliography}{}
\bibitem[Bertone et al.(2004)]{bert04} Bertone, E., Buzzoni, A., Rodriguez-Merino, L.H., \& Chavez, M. 2004, Mem.SAIT, 75, 158
\bibitem[Olive \& Skillman(2004)]{oli04} Olive, K.A. \& Skillman, E.D. 2004, ApJ, 617, 29
\bibitem[Pagel et al.(1992)]{pagel92} Pagel, B.E.J., Simonson, E.A., Terlevich, R.J., \& Edmunds, M.G. 1992, MNRAS, 255, 325
\bibitem[Peimbert \& Torres-Peimbert(1974)]{peim74} Peimbert, M. \& Torres-Peimbert, S. 1974, Ap.J., 193, 327
\bibitem[Rodriguez-Merino et al.(2005)]{rodri05} Rodriguez-Merino, L. H., Chavez, M., Bertone, E., \& Buzzoni, A. 2005, ApJ, 626, 411
\end{thebibliography}
\end{document}